\documentclass{arx}
 
\JournalSubmission   

\usepackage{microtype}                 
\PassOptionsToPackage{warn}{textcomp}  
\usepackage{textcomp}                  
\usepackage{mathptmx}                  
\usepackage{times}                     
\usepackage{tabu}                      
\usepackage{booktabs}
\usepackage{enumerate}
\usepackage{enumitem}
\usepackage{float}
\usepackage{dirtytalk}
\usepackage{csquotes}
\usepackage{amsmath}
\usepackage{amssymb}
\usepackage{subcaption}
\usepackage{bbm}
\usepackage[export]{adjustbox}
\usepackage{todonotes}
\usepackage{siunitx}
\usepackage{mathtools}
\usepackage{url}
\usepackage{algorithm}
\usepackage[noend]{algpseudocode}
\usepackage{dutchcal} 
\usepackage{multirow}

\DeclareMathAlphabet{\mathcal}{OMS}{cmsy}{m}{n}

\newcommand{\N}{\mbox{\rm \hbox{I\kern-.15em\hbox{N}}}}

\newcommand{\rev}[1]{{\textcolor{black}{#1}}}

\algrenewcommand\textproc{}
\algnewcommand{\LineComment}[1]{\vspace{1mm}\State \textit{\small // #1}}

\usepackage[T1]{fontenc}
\usepackage{dfadobe}

\usepackage{cite}  
\BibtexOrBiblatex

\electronicVersion
\PrintedOrElectronic
\usepackage{graphicx}

\title[Neural Face Skinning for Mesh-agnostic Facial Expression Cloning]{Neural Face Skinning for Mesh-agnostic Facial Expression Cloning}

\author[Cha et al.]{
{
    \parbox{\textwidth}{\centering 
        Sihun Cha$^{1}$\orcid{0000-0001-9506-9438},
        Serin Yoon$^{1}$\orcid{0009-0008-7464-9968},
        Kwanggyoon Seo$^{2}$\orcid{0000-0003-0570-4915}, 
        Junyong Noh$^{1}$\orcid{0000-0003-1925-3326}
    }       
}
\\
    {\parbox{\textwidth}{\centering 
        $^1$Visual Media Lab, KAIST, Republic of Korea\\
        $^2$Flawless AI, United States of America 
        }
    }
}
\begin{document}

\teaser{
    \centering
    \includegraphics[width=\linewidth]{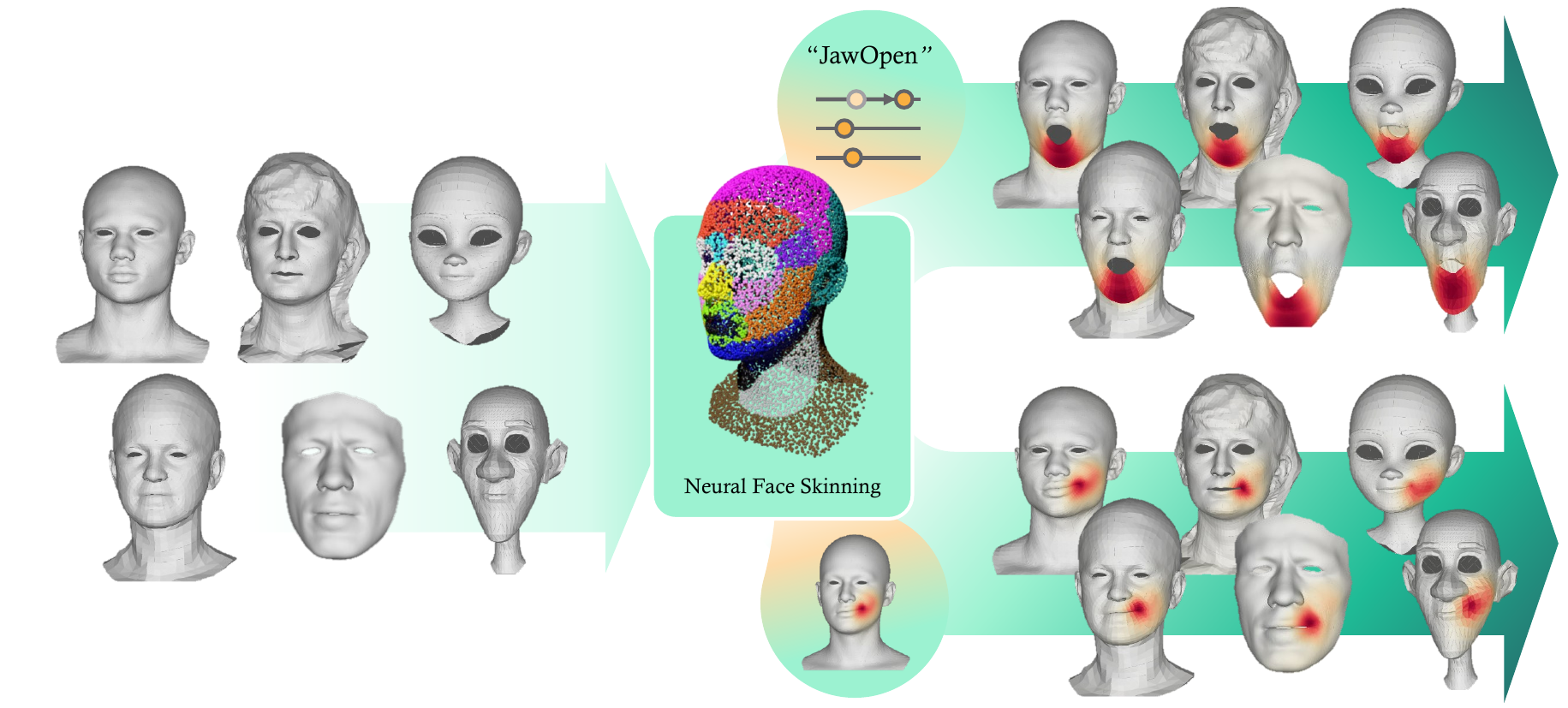}
    \vspace{9mm}
    \caption[teaser]{We present a method that enables direct retargeting between two facial meshes with different shapes and mesh structures. Our method performs well even on facial meshes with proportions that deviate from typical human faces.
    (© Face model: ICT-Facekit~\cite{li2020ictface}, Multiface~\cite{wuu2022multiface}, meryproject.com, VOCASET~\cite{VOCA2019}, BIWI~\cite{fanelli2013biwi}, 2023 AnimSchool)
    }
\label{fig:teaser}
}


\maketitle
\begin{abstract}
Accurately retargeting facial expressions to a face mesh while enabling manipulation is a key challenge in facial animation retargeting. Recent deep-learning methods address this by encoding facial expressions into a global latent code, but they often fail to capture fine-grained details in local regions. While some methods improve local accuracy by transferring deformations locally, this often complicates overall control of the facial expression. To address this, we propose a method that combines the strengths of both global and local deformation models. Our approach enables intuitive control and detailed expression cloning across diverse face meshes, regardless of their underlying structures. The core idea is to localize the influence of the global latent code on the target mesh. Our model learns to predict skinning weights for each vertex of the target face mesh through indirect supervision from predefined segmentation labels. These predicted weights localize the global latent code, enabling precise and region-specific deformations even for meshes with unseen shapes. We supervise the latent code using Facial Action Coding System (FACS)-based blendshapes to ensure interpretability and allow straightforward editing of the generated animation. Through extensive experiments, we demonstrate improved performance over state-of-the-art methods in terms of expression fidelity, deformation transfer accuracy, and adaptability across diverse mesh structures.


\end{abstract}  

\section{Introduction}

Creating natural movements for characters is an important topic in computer graphics. Facial animations are particularly important as they play a key role in communication and emotional expression. Various methods have been developed to create and manipulate natural facial animations for the character faces. Blendshape is one of the most representative methods, providing linear parameterization for face model~\cite{lewis2014practice}. It is widely used because it allows for semantically consistent configurations across various character faces. When the identical blendshape or facial rig is used for different face meshes, it is referred to as \textit{corresponding parameterization}~\cite{lewis2014practice}. 
In this case, the retargeting problem can be solved by transferring the control parameters from one to another. However, the \textit{corresponding parameterization} is not available in general, often leading to the requirement of manual processing. 

The advent of deep learning has made it possible to automate the process of retargeting across various shapes of face meshes, extending the possibility of \textit{corresponding parameterization}~\cite{bouritsas2019neural3dmm,chandran2020semantic,chandran2022shape,groueix20183d,gao2018automatic,jiang2019disentangled,qin2023NFR,ranjan2018coma,tan2018variational}.
Although these methods differ in the network architectures, they share a common approach: encoding the source expression into a latent code and decoding the deformation assumed by the latent code on the target mesh. 
The latent code that represents the expression of the entire face mesh is often referred to as a global code.
It is very useful to compress the deformation of the mesh into an implicit global code as it compactly reduces the control space for animating and manipulating the mesh.
Recent methods further improved this approach by predicting local deformations using global code, such as per-triangle Jacobians~\cite{aigerman2022njf,qin2023NFR} or per-vertex displacements~\cite{chandran2022shape,wang2023zpt}. This eliminated the need for manually defining correspondences between meshes and enabled retargeting across meshes with different structures.
Unfortunately, because the global code lacks the ability to capture the detailed deformations required for each local region, the expression from the source may not be accurately retargeted to the corresponding facial regions of the target.

In contrast to approaches that use the global code, another branch of research aims to retarget facial expression by utilizing local deformations~\cite{brunton2014multilinear,chandran2022local,chandran2024anatomically,joshi2006learning,ma2011optimized,neumann2013sparse,rackovic2021clustering,tena2011interactive,wang2020facial,wu2016anatomically}. While the details of these methods vary, their common idea is to divide the face into multiple regions and transfer the deformation occurring in each region individually. This allows for a higher expressibility of local facial details than the global approaches mentioned above. Because the local deformation methods typically define local regions based on the learned mesh data, their applicability is constrained to a face mesh with the same mesh structure. Additionally, handling local deformations individually often makes it hard to intuitively control the overall facial expression.

\begin{figure}[t]
\begin{center}
\includegraphics[width=\linewidth]{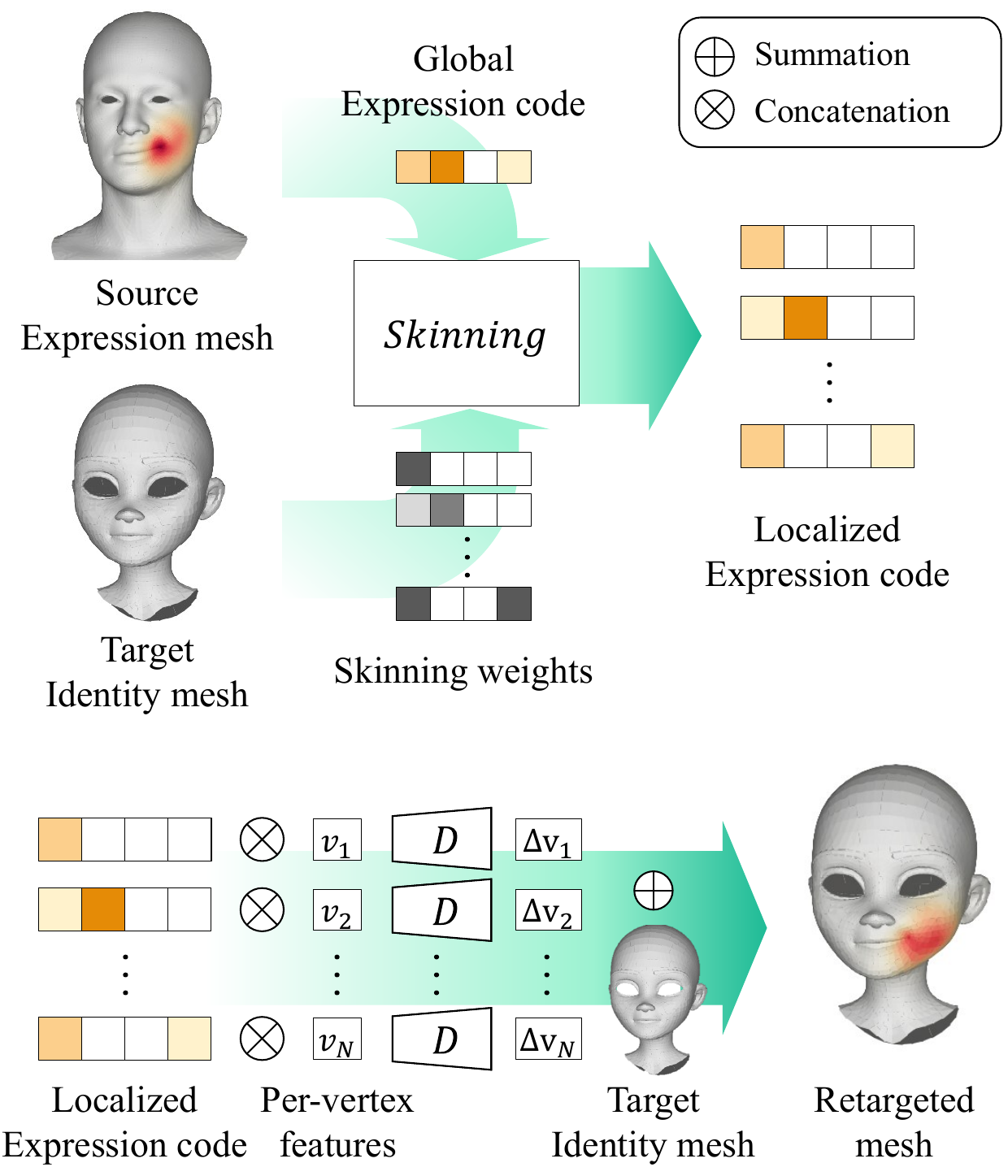}
\end{center}
\caption[Brief illustration of the key idea]{
    Brief illustration of the key idea. Our method localizes the influence of global expression on the local regions in the target identity mesh by utilizing per-vertex skinning weights, enabling precise expression cloning on the local region.
}
\label{fig:inference}
\end{figure}

In this paper, we propose a method to utilize the advantages of both global and local approaches. 
The key idea of our approach is illustrated in Figure~\ref{fig:inference}. 
Our method localizes a global expression code for each vertex based on target mesh geometry. Then the local deformation is predicted using the localized expression code to produce a deformed mesh. We employ a skinning encoder that predicts the per-vertex skinning weight from the target face mesh to localize the global expression code. The skinning weight is then processed with the global expression code to produce a localized expression code for each vertex of the target face mesh. This enables our method to accurately reflect the local geometry of the target face mesh, resulting in precise and realistic retargeting of the source expression, even for meshes with facial proportions that differ from the training data.

In addition to accurate retargeting, the ability to manipulate facial expressions is very crucial. In most deep learning-based methods, the global code is not interpretable, making direct editing of the results challenging. Because it is not feasible for users to manipulate each vertex of the mesh directly to achieve desired outcomes, further editing becomes practical only when the generated mesh is mapped to a controllable format, such as a user-friendly rig or blendshapes, through an inverse processing procedure~\cite{villanueva2022voice2face}. To tackle this, we construct the expression code based on the Facial Action Coding System (FACS)~\cite{ekman1978facial}. This allows our method to be interpretable, facilitating manipulation and adjustment of the generated output.
The experiments demonstrate that our approach effectively allows for expression cloning and follow-up manipulation while accurately reflecting the local geometry information of the target face mesh. We show that our method outperforms the baseline methods on the quality of retargeting and inverse rigging. Furthermore, our method performs robustly even on face meshes with different proportions from those of the meshes used as the training data.

\section{Related Work}

A parametric model is a commonly used method to control face meshes, offering a simple and effective way to represent facial expressions with a small number of parameters~\cite{parke1991control}. There are various types of parametric facial models, such as blendshape and 3D Morphable Models (3DMM). 
3DMM is often constructed from the captured data with PCA, producing a linear model with an orthogonal basis~\cite{blanz19993dmm,choe2006analysis,li2017flame,paysan2009bfm,vlasic2006face}. Because the orthogonality of the basis does not guarantee to match the shape semantic of the facial expression, the interpretability is low and it is hard for users to edit the model. 
The blendshape is also a linear model that provides a semantic basis where each basis corresponds to an individual facial expression~\cite{lewis2014practice}. The type of blendshape varies depending on the blending target. When the entire face shape is linearly blended~\cite{parke1972computer,parke1991control}, it is referred to as a whole-face blendshape. In the case of delta blendshape~\cite{bergeron19873}, the offset for the expressive face is blended linearly with the neutral face. While these two types of blendshape handle the expression globally upon the entire face, local blendshape~\cite{kleiser1989fast} divides the face into several regions, blending each segmented face region individually to produce a wider range of expressions.
We recommend the survey works of \textit{Lewis and Anjyo}~\cite{lewis2014practice} and \textit{Egger et al.}~\cite{egger20203d} for a further detailed review of the relevant research.

Because a parametric model is defined based on the mesh structure, they are bound to a specific face mesh.
A simple way to solve this is to establish correspondences between source and target meshes and transfer per-vertex displacements~\cite{noh2001expression} or deformation gradients~\cite{sumner2004deformation}. These approaches were later improved by utilizing Radial Basis Functions~\cite{bickel2007multi,orvalho2008transferring,ribera2017facial}, retargeting motion in the velocity domain~\cite{seol2012spacetime}, or leveraging segments for local retargeting~\cite{liu2011framework}.
These methods can be easily applied to the blendshape that handles the face globally. However, it is not clear how to extend these approaches to handle local blendshapes as they require defining local segments in addition to fitting a global shape.

The emergence of deep learning has significantly expanded the possibilities for facial model parameterization. This has been achieved by encoding mesh deformation into latent space using various network architectures. These architectures include Multi-Layer Perceptrons~\cite{aigerman2022njf,groueix20183d,tan2018variational,qin2023NFR,wang2023zpt}, 2D Convolutional Networks~\cite{bagautdinov2018modeling}, Graph Convolutional Networks (GCN)~\cite{bouritsas2019neural3dmm,gao2018automatic,hanocka2019meshcnn,ranjan2018coma}, and Transformers~\cite{chandran2022shape}. These studies can be broadly categorized into two main types: global approaches and local approaches.

Early studies of the global approach represent the deformation of a mesh using a single global code.
Ranjan et al.~\cite{ranjan2018coma} use GCN to learn the global latent representation of facial expressions.
Bouritsas et al.~\cite{bouritsas2019neural3dmm} further improved this by employing a spiral convolutional network, a variant of GCN. 
Groueix et al.~\cite{groueix20183d} utilize a template mesh along with the global shape code which encapsulates the deformation. Tan et al.~\cite{tan2018variational} use variational autoencoders to learn a latent space for the deformation. Gao et al.~\cite{gao2018automatic} employed a GAN-based model with a cyclic loss to learn the global latent representation for mesh deformation using unpaired data, enabling deformation transfer to a target mesh. 
Some approaches~\cite{chandran2020semantic,jiang2019disentangled} separate the global code into identity and expression components. This helped the model to disentangle the information, improving the quality of expression retargeting. 
Recent methods enable the retargeting on face meshes with arbitrary structures via learning the field of local deformation, such as per-triangle Jacobian~\cite{aigerman2022njf} or per-vertex displacement~\cite{chandran2022shape,wang2023zpt}. Qin et al.~\cite{qin2023NFR} further improved this by constructing the global code based on FACS, providing easy control for creating and editing facial expressions. Building on these approaches, we localize the influence of the global code to improve retargeting accuracy while enabling intuitive control.

The local approach utilizes a local model to improve expressibility by discarding undesired spatial correlation biases~\cite{brunton2014multilinear,wang2020facial,chandran2022local,chandran2024anatomically}. 
Brunton et al.~\cite{brunton2014multilinear} utilize multilinear models based on wavelet coefficients to effectively capture local details of facial expressions. However, these models sometimes produce an unnatural expression as a whole.
To address this, Bagautdinov et al.~\cite{bagautdinov2018modeling} utilize a 2D CNN to capture both global and local deformations by projecting the mesh deformation into the UV space. Wang et al.~\cite{wang2020facial} utilize the global and local multilinear models simultaneously, to produce natural expression as a whole.
While the details of these methods vary, their common idea is to divide the face into multiple regions and retarget the deformation in each region individually. 


\begin{figure*}[t]
\begin{center}
\includegraphics[width=\linewidth]{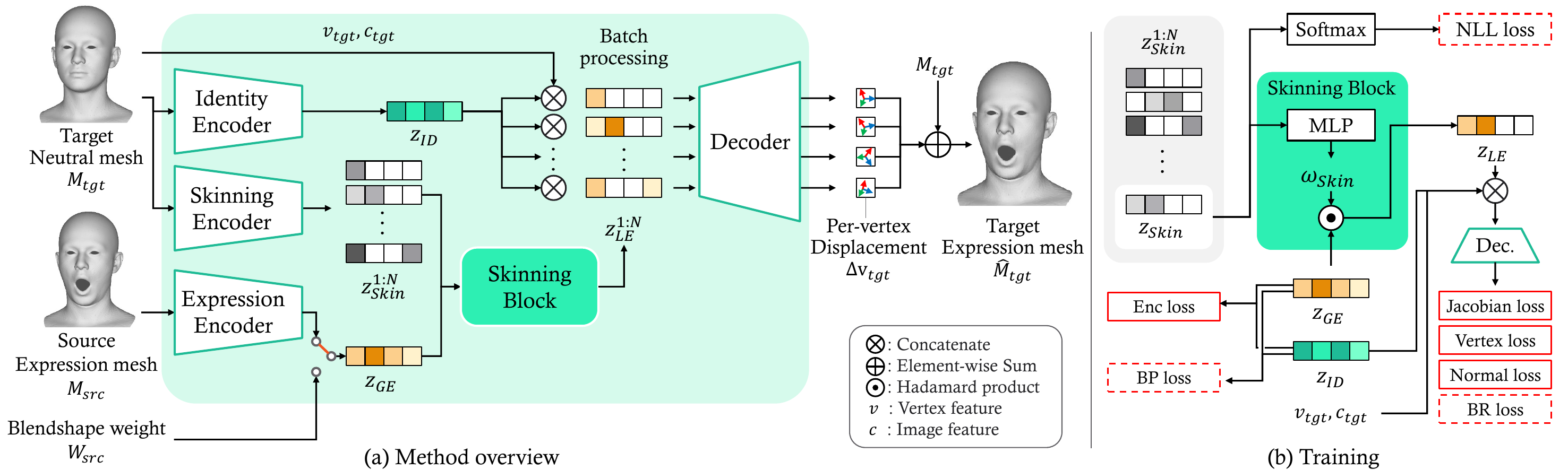}
\end{center}
\caption[Method overview]{
    Method overview. Illustration of the data flow at inference (a) and training (b). For simplicity, the encoders are omitted from (b) and the dotted red box indicates the losses that are exclusively applied to the ICT data.
}
\vspace{-2mm}
\label{fig:method_overview}
\end{figure*}

The strengths and weaknesses of both global and local approaches often involve a trade-off between intuitive control and preservation of local details. We aim to mitigate the limitations of each approach while leveraging their advantages. Our method predicts the local influence of the global expression code on the target mesh through skinning weights. This enables expression cloning in localized regions of the face, combining the benefits of both global and local methods for natural and precise retargeting. To obtain appropriate skinning weights for any mesh structure, we use segmentation labels as indirect supervision during network training. Following Qin et al.~\cite{qin2023NFR}, we guide the global expression code to function as FACS-based blendshape weights, allowing users to intuitively edit facial expressions.

\section{Method}


In this section, we present the flow of our method, along with the network architecture and training strategy. As shown in Figure~\ref{fig:method_overview} (a), our approach allows for retargeting expressions from a given source face mesh to a neutral target mesh, as well as animating based on user-provided blendshape weights. In the retargeting setting, our model takes the source expression mesh $M_{src}$ and the target neutral mesh $M_{tgt}$ as inputs. For animation using blendshapes, the model uses the blendshape weight $W_{src}$ and the target neutral mesh $M_{tgt}$ as inputs. 
In both cases, the output is the displacement $\Delta \text{v}_{tgt}$ for each vertex of $M_{tgt}$, which is added to the original vertex positions of $M_{tgt}$ to produce the deformed mesh.


\subsection{Architecture}

Our method utilizes three encoders and one decoder. The three encoders are the identity encoder, expression encoder, and skinning encoder. The identity encoder encodes identity information from $M_{tgt}$ into a global identity code $z_{ID} \in \mathbb{R}^{128}$, while the expression encoder encodes facial expression information from $M_{src}$ into global expression code $z_{GE} \in \mathbb{R}^{128}$. The skinning encoder initially predicts the skinning feature $z_{Skin} \in \mathbb{R}^{L}$ for each vertex of $M_{tgt}$ where $L$ represents the number of segmentation labels. Given $z_{Skin}$ and $z_{GE}$, the skinning block outputs the localized expression code $z_{LE} \in \mathbb{R}^{128}$.

\begin{figure}[t]
\begin{center}
\includegraphics[width=\linewidth]{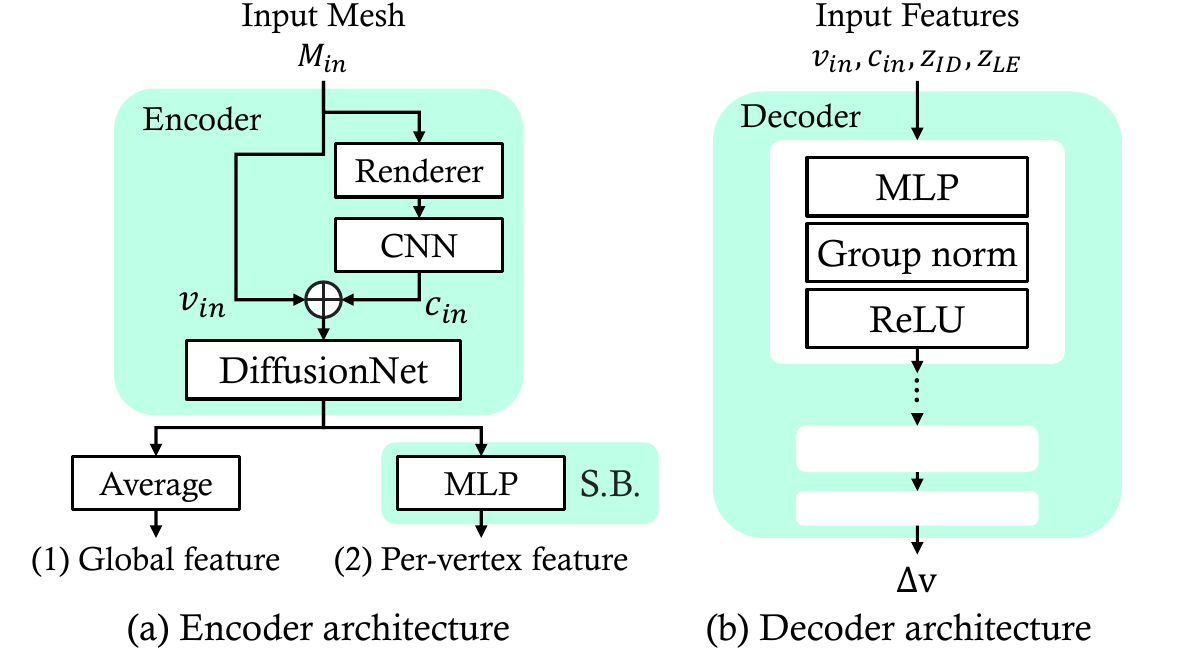}
\end{center}
\caption[Illustration of the encoder and decoder architectures.]{
    Illustration of the encoder (a) and decoder (b) architectures.
    S.B. indicates the skinning block.
}
\label{fig:architecture}
\end{figure}

All encoders contain a CNN and DiffusionNet~\cite{sharp2022diffusionnet}, following the approach of Qin et al.~\cite{qin2023NFR}. The skinning encoder additionally has a skinning block, which consists of simple MLP. When the input mesh $M_{in}$ is fed into the encoders, the face mesh is first rendered from a frontal view, and the CNN extracts the image feature $c_{in}$. This feature $c_{in} \in \mathbb{R}^{128}$ is concatenated with the vertex feature $v_{in} \in \mathbb{R}^{6}$ of $M_{in}$ and passed into DiffusionNet. 
We use $c_{in}$ to enhance the network's robustness to alignment differences between a given face mesh and a learned face mesh, as noted in Qin et al.~\cite{qin2023NFR}.
The vertex feature is the concatenation of the vertex position and the vertex normal.
While all three encoders share the same input representation, they differ in how the DiffusionNet output is processed. For the identity and expression encoders, the outputs from all vertices are averaged to produce a single global code. In contrast, the skinning encoder's output is used directly without averaging as shown in Figure~\ref{fig:architecture} (a).


The MLP in the skinning block takes $z_{Skin}$ as input and outputs skinning weights $\omega_{Skin} \in \mathbb{R}^{128}$ where $128$ represents the number of blendshapes. By applying the Hadamard product to $\omega_{Skin}$ and $z_{GE}$ the localized expression code ${z}_{LE} \in \mathbb{R}^{128}$ associated with each vertex is produced. The decoder consists of an eight-layer MLP, where the output of each layer, except for the final one, is followed by group normalization and ReLU activation. The decoder takes the concatenation of $v_{tgt}$, $c_{tgt}$, $z_{ID}$ and $z_{LE}$ as input and outputs the per-vertex displacement $\Delta \text{v}_{tgt}$ as shown in Figure~\ref{fig:architecture} (b). This displacement is then added to the vertex positions of $M_{tgt}$ to produce the deformed mesh. 
The decoder processes each vertex independently and supports batch processing, allowing it to handle all vertices in a mesh at once without being influenced by neighboring vertices.

\subsection{Dataset}

For the training data, we utilized the ICT-Facekit~\cite{li2020ictface} (ICT), a FACS-based parametric face model. ICT provides identity and expression blendshapes, which we used to supervise both the identity and expression encoders. Specifically, we ensured that the global expression code, $z_{GE}$, mimics the ICT expression blendshape coefficients, allowing for an interpretable global code that facilitates intuitive facial expression manipulation. The ICT blendshape is based on a delta blendshape formulation as follows:
\begin{equation}
\text{v}^{i}_{0} + \sum^{J}_{j=1} w^{i}_{j}(\text{v}^{i}_{j} - \text{v}^{i}_{0}) + \sum^{K}_{k=1} w^{i}_{k}(\text{v}^{i}_{k} - \text{v}^{i}_{0})
\end{equation}
where $\text{v}^{i}_{0} \in \mathbb{R}^3$, $\text{v}^{i}_{j} \in \mathbb{R}^3$, and $\text{v}^{i}_{k} \in \mathbb{R}^3$ are the $i$-th vertex of `neutral' face, $j$-th identity blendshape, and $k$-th expression blendshape, respectively. $w_j$ and $w_k$ are the corresponding coefficients for identity blendshape and expression blendshape, respectively. $J$ represents the number of identity blendshapes and $K$ represents the number of expressions which are $100$ and $53$ in ICT, respectively. 

We generated synthetic data by randomly sampling expression coefficients $w_k$ from a uniform distribution. Additionally, we used data where expression coefficients were sampled in a one-hot manner to ensure that the model could fully learn extreme expressions represented by individual blendshapes.
We sampled $111$ identity coefficients $w_j$ from the normal distribution to generate a variety of identities in the dataset. Specifically, the $100$ identity coefficients publicly released by Qin et al.~\cite{qin2023NFR} were used for training, and $10$ identities were sampled for testing and $1$ identity was sampled for validation. 

In addition to the synthetic data, we incorporated the Multiface~\cite{wuu2022multiface} dataset for further training. Multiface is a dataset created by capturing the range of motion (ROM) for $13$ different identities.
We split the identities as $8:1:1$ each for training, validation, and testing, following Qin et al.~\cite{qin2023NFR}.
This real scan data helps the model learn realistic facial expressions and movements that are difficult to capture using ICT blendshapes alone. We split the training, validation, and test sets based on identity for both Multiface and ICT to ensure diverse learning. 
Additionally, all face meshes used in training were standardized by removing the eyeballs and mouth sockets, and the precomputed features were calculated after this standardization process.

\subsection{Skinning prediction}
A key aspect of our method is predicting the skinning weights that capture the relationship between facial regions and the global expression code. 
Skinning, typically used in joint-based mesh deformation, determines how a joint’s movement affects the mesh's vertices. A well-known example is Linear Blend Skinning (LBS)~\cite{magnenat1988LBS}, where the influence of each joint is usually limited to a local region instead of affecting the entire mesh. Inspired by this, we propose a method that localizes the global expression code through skinning weights, resulting in accurate region-specific deformations.

Unlike joint-based methods, blendshape face models do not utilize pre-defined skinning weights. To address this, we propose a strategy where the skinning encoder implicitly learns to predict skinning weights for the given mesh using supervision from segmentation labels. This enables our method to localize the influence of $z_{GE}$ on different facial regions, resulting in accurate and expressive deformations. To supervise the skinning encoder using segmentation labels, we created a segmentation map that divides the face into several regions based on the facial muscle group~\cite{winslow2015classic}. We then apply a negative log-likelihood loss to the output of the skinning encoder as follows:
\begin{equation} \label{eq:L_nll}
L_{nll} = - \frac{1}{N}\sum^{N}_{i=1} \left(y^{i} \log(z^{i}_{Skin}) + (1 - y^{i}) \log(1 - z^{i}_{Skin})\right)
\end{equation}
where $N$ denotes the total number of vertices, and $y$ represents the ground truth labels.

We do not force $z_{Skin}$ to take a strict one-hot form, allowing for some degrees of spatial correlation to remain during training. In other words, rather than limiting each vertex to belong exclusively to a single region, the model allows for a soft association between vertex and multiple regions. As a result, our method is capable of handling both global and local deformations.

\begin{figure}[t]
\begin{center}
\includegraphics[width=\linewidth]{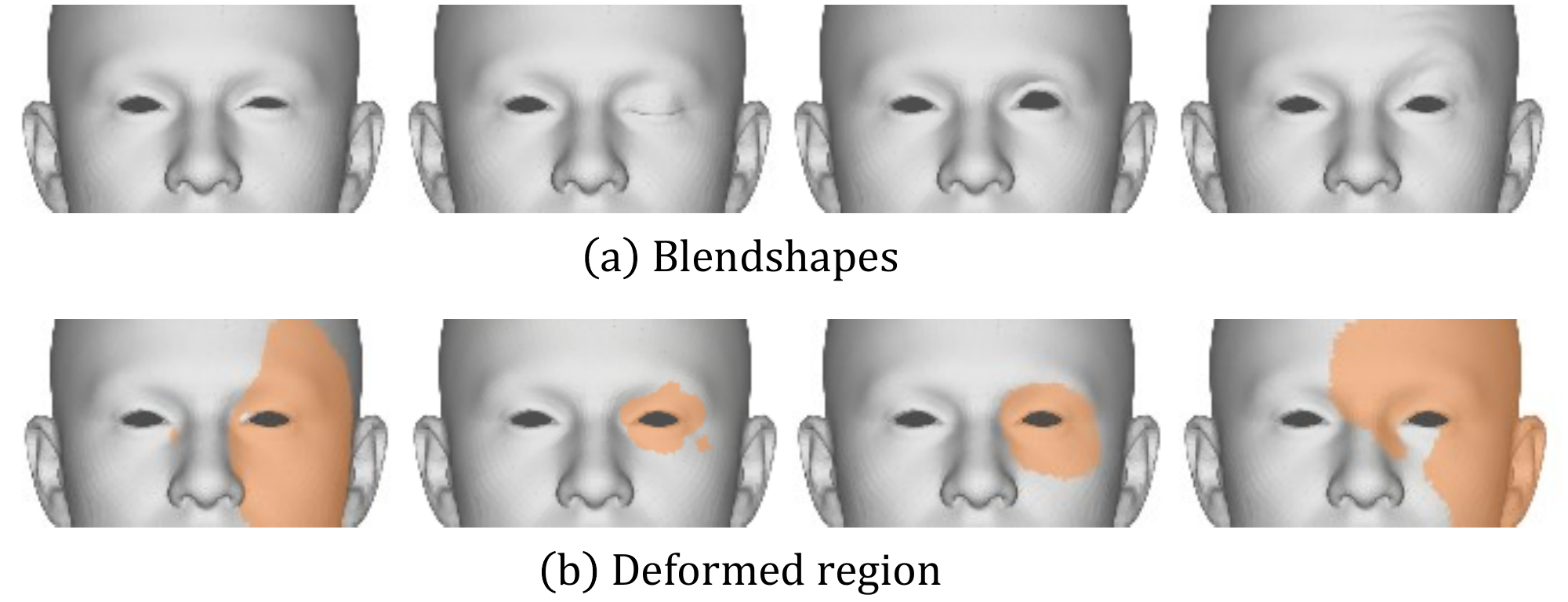}
\end{center}
\caption[Overlaping regions between blendshape]{
    Overlaping regions between blendshape. The ICT blendshapes for `EyeSquintRight', `EyeBlinkRight', `EyeLookUpRight', and `BrowOuterUpRight' (a) and the corresponding deformed region (b).
}
\label{fig:blendshape_region}
\end{figure}

As shown in Figure~\ref{fig:blendshape_region}, the areas influenced by different blendshapes often overlap, even when the semantics of the movements differ. More cases can be found in the supplementary material. Therefore, after segmenting the face into regions, we trained the network to map these regions to skinning weights, by implicitly learning the correlation between the blendshape movements and the facial region. The effectiveness of this approach is demonstrated in Section~\ref{sec:exp_quality}.

Indirect supervision using segmentation aims to regularize the skinning encoder by ensuring consistent skinning weights for corresponding face regions across different face shapes. While the encoder can converge without this supervision, the estimated weights may lack the desired consistency. Additionally, without supervision, the encoder showed reduced generalization to unseen meshes as shown in Figure~\ref{fig:application_unseen}. 
Supervising only a subset of the training data proved sufficient for the network to learn facial semantics and generalize to diverse face shapes and structures. For related results, please refer to the supplementary material.

\subsection{Loss function}

In the training process, we use several loss functions: decoder loss, encoder loss, negative log-likelihood loss based on segmentation, and losses based on ICT blendshapes. 
The decoder loss is based on the L2 loss between the predicted deformed mesh and the ground truth mesh, comprising vertex loss ($L_v = || v-v_{GT}||^2$), Jacobian loss ($L_g = || g-g_{GT}||^2$), and normal loss ($L_n = || n-n_{GT}||^2$), where $v$, $g$, and $n$ represent the vertex position, deformation Jacobian, and vertex normal of the predicted mesh, respectively. Subscript $GT$ refers to the ground truth. The loss can be expressed as follows:
\begin{equation}
L_{dec} = \lambda_v L_v + \lambda_n L_n + \lambda_g L_g
\end{equation}
where $\lambda_v$, $\lambda_n$ and $\lambda_g$ are set to $10$, $1$, and $1$, respectively.

The encoder loss calculates the L2 loss between the encoder predictions and the ground truth ICT blendshape coefficients. 
ICT provides identity blendshapes with $100$ bases and expression blendshapes with $53$ bases. Because both $z_{ID}$ and $z_{GE}$ are represented as $128$-dimensional vectors, the extra dimensions beyond these blendshape bases are regularized to approach to zero. 
The losses for the identity and expression encoders are defined as follows:
\begin{equation}
\begin{split}
L_{ID} = || z^{id}_{ID} - z^{id}_{GT}||^2 + ||z^{ext}_{ID}||^2, \\
L_{Exp} = || z^{exp}_{GE} - z^{exp}_{GT}||^2 + ||z^{ext}_{GE}||^2.
\end{split}
\end{equation}
where \textit{id} and \textit{exp} refers to the identity and expression blendshape dimensions, respectively, and \textit{ext} refers to the extra dimensions.


For non-ICT data, we apply a regularization loss ($L_{reg}$) as proposed in NFR~\cite{qin2023NFR}. This regularization loss ensures that the predicted parameters remain within the range [0, 1], and it is defined as follows:
\begin{equation} \label{eq:L_r}
L_{reg} = 
    \begin{cases} 
    -z, & z < 0 \\
    0, & 0 \leq z \leq 1 \\
    z - 1. & z > 1 
    \end{cases}
\end{equation}
where $z$ represents both the expression and identity codes.

Summarizing the loss based on the training data yields the following:
\begin{equation}
L_{enc} = 
\begin{cases} 
L_{Exp} + L_{ID}, & \text{if ICT} \\
L_{reg}. & \text{else} 
\end{cases}
\end{equation}


In addition to the encoder loss, we introduce two additional losses based on the ICT blendshape ($L_{BP}$, $L_{BR}$) to align the latent code from the expression encoder with the blendshape coefficients and guide the decoder to behave similarly to the ICT blendshape basis. These losses are applied exclusively when the training data is obtained from ICT. 
Considering the vertex as an end-effector of the blendshape system, applying the losses to minimize its deviation helps the network better align with the blendshape system compared to using only an encoder loss. Table~\ref{tab:ablation} and Table~\ref{tab:ablation_inv_rig} show that the blendshape-based losses improve the quality of both the expression encoder and decoder outputs.

The blendshape projection loss ($L_{BP}$) minimizes the L2 distance between the expression face reconstructed by multiplying $z_{GE}$ with the ICT blendshape basis and the expression mesh generated by multiplying the ground truth ICT blendshape coefficients with the same basis. This encourages the prediction of the expression encoder to behave like ICT blendshapes.
\begin{equation}
    L_{BP} = || z_{GE}^{1:53} \cdot B - z_{GT} \cdot B ||^2
\end{equation}
where $B$ represents the ICT blendshape basis.

The blendshape reconstruction loss ($L_{BR}$) minimizes the L2 distance between the expression face generated by the ICT blendshape and the predicted expression mesh from the decoder. The first $53$ dimensions of $z_{GE}$ are multiplied by $B$ to generate an expression mesh, while the decoder uses the entire $z_{GE}$ to produce its output. This loss ensures that the decoder learns to behave similarly to the ICT blendshape basis.
\begin{equation}
    L_{BR} = || \textbf{sg}(z_{GE}^{1:53}) \cdot B - D(\Psi(\textbf{sg}(z_{GE}), z_{Skin})) ||^2
\end{equation}
where $\textbf{sg}$ represents the stop-gradient operation, $\Psi$ represents the skinning block, and $D$ is the decoder. For simplicity, additional inputs to the decoder are omitted from the equation. We apply the stop-gradient operation to $z_{GE}$ to prevent the gradient from propagating back to the expression encoder, ensuring that the loss affects only the decoder.

The negative log-likelihood loss in Equation~\ref{eq:L_nll} is applied only when the mesh in the training data corresponds to ICT. For other meshes, the training process converges towards minimizing the remaining losses.
The overall loss is defined as follows:
\begin{equation}
L_{total} = \begin{cases} L_{dec} + L_{enc} + L_{BP} + L_{BR} + L_{nll}, & \text{if ICT} \\
L_{dec} + L_{enc}. & \text{else} \end{cases}
\end{equation}

\section{Experiments}

In this section, we outline the experiments carried out to evaluate the effectiveness of our method. To assess the expression fidelity of our method, we compared the quality of facial expression retargeting results with that of the results from \textit{Qin et al.} (NFR)\cite{qin2023NFR} and \textit{Wang et al.} (ZPT)\cite{wang2023zpt}, both of which handle deformation transfer based on global codes and can be applied to arbitrary mesh structures. 
Although \textit{Chandran et al.}~\cite{chandran2022shape} is also independent of specific mesh structures, it does not use global codes. Instead, the model directly predicts delta values of the vertex positions from the target and expression meshes, making it less suitable for tasks that require editable or controllable facial movements.
Therefore, we focused on methods that offer flexibility in control for comparison. We trained all models with ICT and Multiface data for fair comparison.
For ZPT, which requires corresponding pose codes for a deformed mesh, we used ICT blendshape coefficients as the pose codes for ICT meshes. For Multiface which does not provide ICT blendshape coefficients for each expression face mesh, we used the predictions from our pre-trained expression encoder as pseudo ground truth, which were then used to train ZPT.

Our study builds upon several aspects of the experimental setup used by NFR, except for the proposed skinning encoder and blendshape-based loss functions ($L_{BP}$, $L_{BR}$). By keeping these elements consistent, we were able to isolate and clearly assess the impact of our proposed changes. 
Additionally, we conducted an ablation study to validate the effectiveness of our design choices. Further details on the implementation and the learned deformations for each expression code dimension can be found in the supplementary material.

\subsection{Expression quality}\label{sec:exp_quality}

\begin{figure*}[t]
\begin{center}
\includegraphics[width=\linewidth]{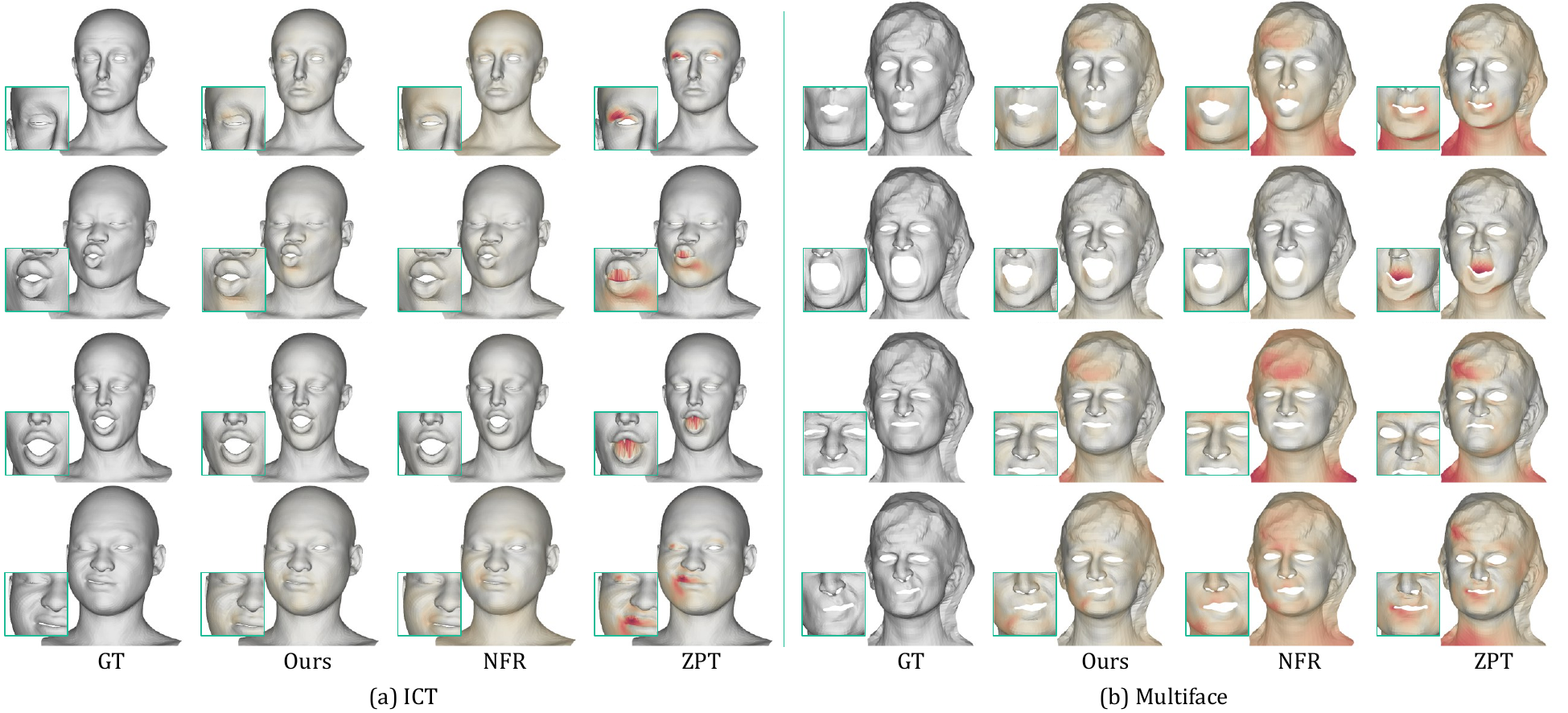}
\end{center}
\caption[Visual comparison of expression quality produced by our method and comparative methods]{
    Visual comparison of expression quality produced by our method and comparative methods.
     The MSE between the GT and the predicted face mesh is colored using a yellow-orange-red color map (YlOrRd).
}

\label{fig:results}
\end{figure*}

\begin{figure*}[t]
\begin{center}
\includegraphics[width=\linewidth]{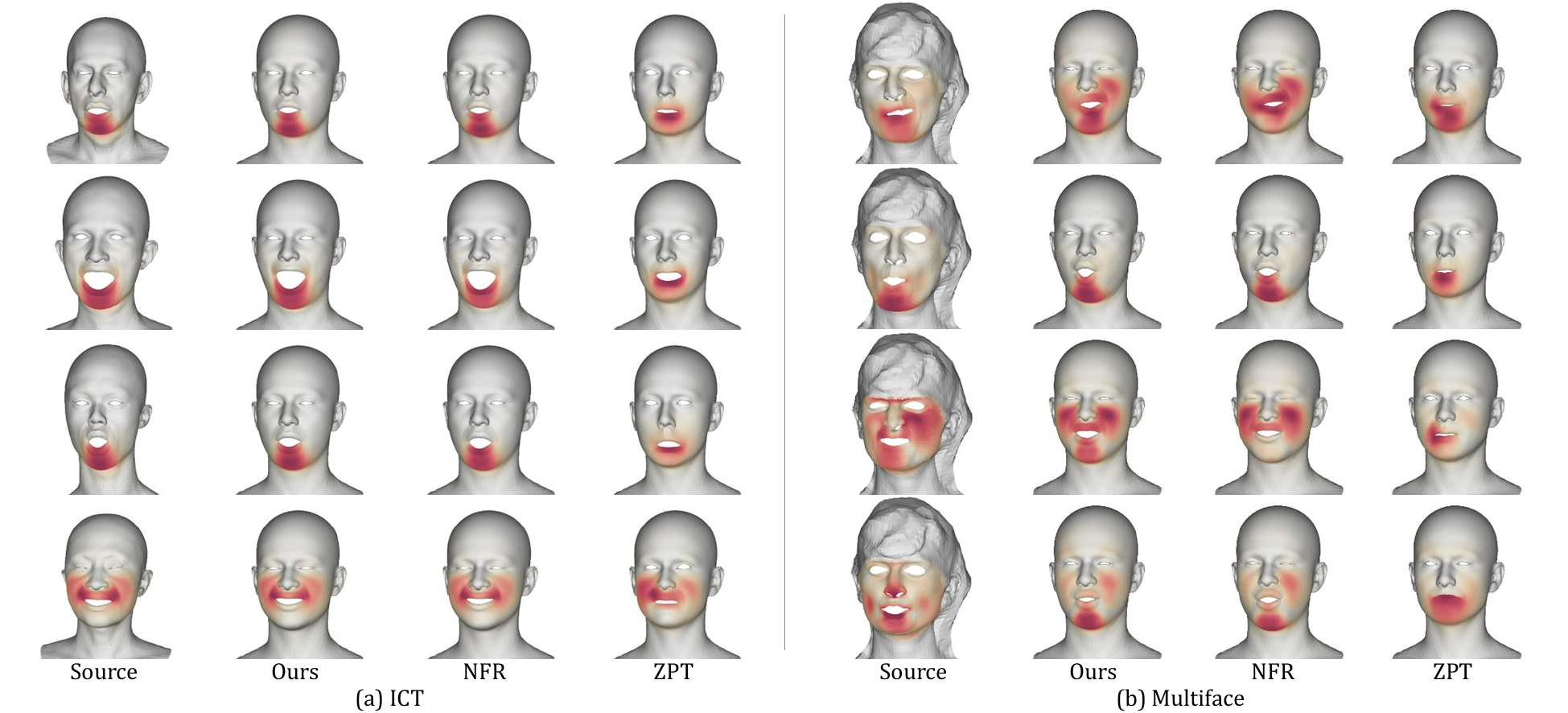}
\end{center}
\caption[Visual comparison of inverse rigging quality produced by our method and comparative methods]{
    Visual comparison of inverse rigging quality produced by our method and comparative methods.
     The expression codes were predicted from the source face mesh and were used to reconstruct a face mesh using the ICT blendshape. The incurred deformation on the face is colored using a yellow-orange-red color map (YlOrRd).
}
\label{fig:results_invrig}
\end{figure*}

To evaluate the expression quality, we conducted a self-retargeting task in which the model retargets a face mesh with an expression to the same mesh in a neutral expression.  We measured the Mean Squared Error (MSE) between the ground truth face mesh and the reconstructed face mesh vertices to quantify the expression quality. For a fair comparison with ZPT, we followed the optimization and test-time training settings outlined in the paper for each identity face mesh in the test data. In case of NFR, the mesh's global mean often shifted or the mesh scale often changed when solving the Poisson equation. To account for this, we aligned the reconstructed mesh with the ground truth using Procrustes analysis before calculating MSE. This minimized errors caused by shifting and ensured that the MSE focused on the deformation errors. The results are shown in Table~\ref{tab:exp_retarget} and Figure~\ref{fig:results} visualize the quantitative errors.

NFR caused subtle but unintended deformations in the mesh, such as slight thinning or thickening of the overall volume, in addition to the shifting artifact. ZPT produced artifacts in regions with conflicting movements, such as when the mouth opens or the eyes blink. In contrast, our method did not produce any of these artifacts and outperformed the comparative approaches.

\begin{table*}[ht]
\caption{
    Quantitative comparison of expression quality on each face segment produced by our method and comparative methods.
}
\centering
\resizebox{\linewidth}{!}{
\begin{tabular}{lcccccccccccccccccccccc}
    \toprule 
    \multirow{2}{*}{Method} & \multicolumn{19}{c}{MSE $\downarrow (\times 10^{-4}mm)$ } \\ 
    \cline{2-22}
         & 01 & 02 & 03 & 04 & 05 & 06 & 07 & 08 & 09 & 10 & 11 & 12 & 13 & 14 & 15 & 16 & 17 & 18 & 19 & 20 & \textbf{Avg} \\
    \midrule
    ZPT & 0.533 & 0.389 & 2.079 & 2.059 & 0.137 & 0.134 & 0.125 & 1.421 & 1.409 & 1.425 & 1.535 & 1.649 & 10.942 & 233.510 & 5.744 & 0.040 & 1.197 & 0.246 & 0.211 & 0.072 & \textbf{13.219} \\

    NFR & 4.389 & 3.233 & 2.545 & 2.218 & 2.893 & 2.075 & 2.226 & 1.639 & 2.298 & 3.218 & 2.613 & 2.853 & 4.584 & 5.657 & 5.106 & 4.272 & 3.913 & 2.468 & 2.869 & 5.861 & \textbf{3.346} \\
    
    Ours & 0.387 & 1.230 & 1.361 & 1.263 & 1.436 & 0.654 & 0.558 & 1.578 & 3.330 & 2.483 & 2.021 & 2.037 & 5.410 & 6.055 & 3.999 & 0.204 & 0.964 & 0.378 & 0.409 & 0.341 & \textbf{1.805} \\

    \bottomrule
\end{tabular}\label{tab:mse_seg}
}
\end{table*}

\begin{figure}[t]
\begin{center}
\includegraphics[width=\linewidth]{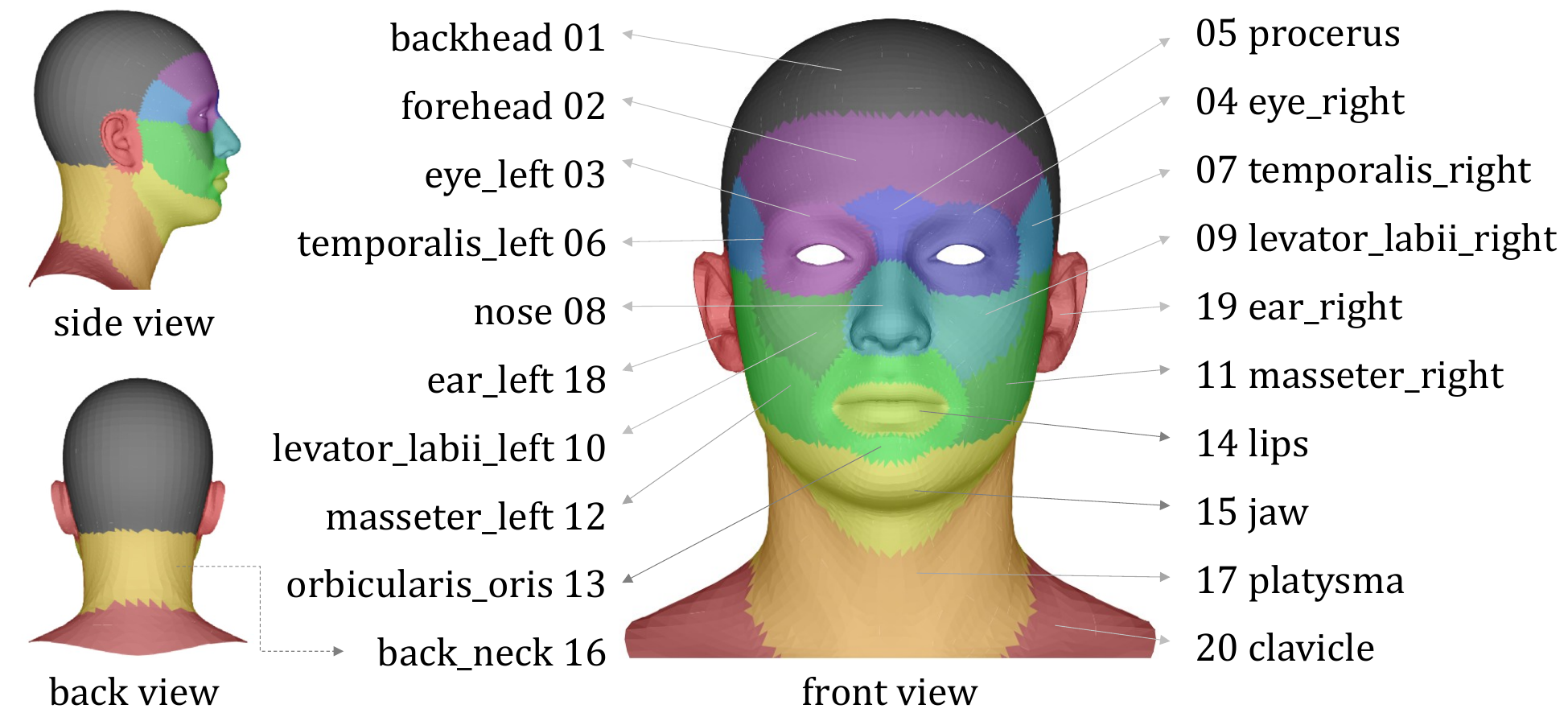}
\end{center}
\caption[Segment label]{
    Segment label. Visualization of 20 segments on the ICT model and their corresponding label.
}
\label{fig:segment_description}
\end{figure}

\begin{figure}[t]
\begin{center}
\includegraphics[width=\linewidth]{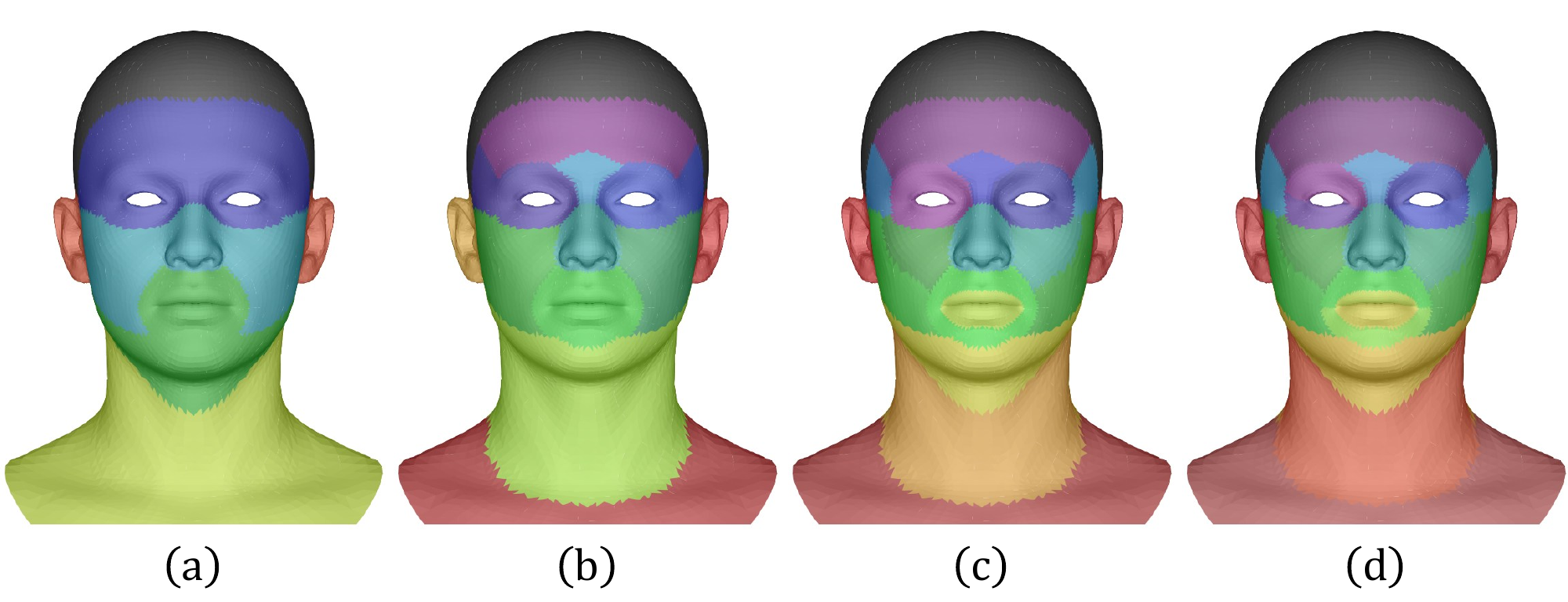}
\end{center}
\caption[Segment variation]{
    Segment variation. Visualization of various segments used for the training (a) 6 segments, (b) 14 segments (c) 20 segments, and (d) 24 segments.
}
\label{fig:segment_variation}
\end{figure}

\begin{table}[h]
\caption[Quantitative comparison of expression quality across different methods]{
    Quantitative comparison of expression quality produced by our method and comparative methods. The best score is indicated in \textbf{bold}.
}
\centering
\begin{tabular}{lcc}
\toprule 
\multirow{2}{*}{Method} & \multicolumn{2}{c}{MSE $\downarrow (\times 10^{-4}mm)$ }   \\ 
\cline{2-3}
              & ICT             & Multiface        \\ 
\cline{1-3}
NFR           & 0.3514          & 5.2257           \\
ZPT           & 2.3327          & 6.6935           \\
Ours          & \textbf{0.2480} & \textbf{4.2856}  \\
\bottomrule
\vspace{-2mm}
\end{tabular}\label{tab:exp_retarget}

\end{table}

The ICT face model includes the full head and part of the torso, leaving relatively few vertices responsible for facial movements. Because regions like the neck and clavicle remain mostly static even when the mouth and eyes move, calculating MSE across the entire mesh can bias the error toward these static regions and underestimate errors in regions with facial movement.
To address this, we calculated the MSE for each predefined face segment and averaged these values for the overall error. The face segment is illustrated in Figure~\ref{fig:segment_description}, and the MSE for each segment, along with the overall average, is presented in Table~\ref{tab:mse_seg}.

While ZPT produced the lowest error in static regions (e.g., 16: backneck, 20: clavicle), however, it produced the highest error in dynamic regions (e.g., 14: lips, 15: orbicularis oris). In contrast, NFR produced relatively consistent errors across both static and dynamic regions. Our method performed similarly to ZPT in static regions while matching NFR's accuracy in dynamic regions. This suggests that our model accurately captures movements where needed and minimizes unnecessary deformation in static areas. Consequently, our method achieved the lowest overall error computed by averaging the segment-based MSE values.



\subsection{Inverse rigging}\label{sec:inv_rig} 
To validate if our constructed expression code conforms to the grammar of the FACS-based ICT blendshape model, we conducted experiments on inverse rigging. Inverse rigging aims to estimate the optimal rig parameters that deform a mesh to the shape of the desired geometry. 
To measure the accuracy of inverse rigging, the expression codes were predicted from the source face mesh with expressions and applied to the ICT blendshape bases. The MSE between the reconstructed face mesh and the source face mesh was then computed.
We only used the ICT data for the quantitative comparison as Multiface does not use a blendshape representation. Because ZPT does not directly predict the expression code, we obtained it through an optimization process by minimizing the error between the deformed output mesh and the source mesh while keeping the model parameters fixed. Table~\ref{tab:inv_rig} shows that our method outperformed the comparative approaches.

Figure~\ref{fig:results_invrig} shows a visual comparison of our results with those produced from comparative methods for both ICT and Multiface. 
NFR performed poorly on Multiface, sometimes producing movements not present in the source. For example, in the first and the third rows of Figure~\ref{fig:results_invrig} (b), the source face has an expression of mouth moving to the left and squinting, respectively. However, the inverse rigging results produced by NFR added unintended expressions, such as eyes being closed. This indicates that some expressions are entangled in the expression code produced by the NFR's encoder. In contrast, the results from our method accurately captured the semantics of the expression without the entanglement of the unintended expressions. This demonstrates the effectiveness of our approach in encoding the facial expressions precisely.

\begin{table}[h]
\caption{
    Quantitative comparison of inverse rigging quality produced by our method and comparative methods. The best score is indicated in \textbf{bold}.
}
\centering
\begin{tabular}{lc}
    \toprule 
    Method & MSE $\downarrow (\times 10^{-4}mm)$ \\ \midrule
    NFR   & 0.5029 $\pm$ 0.3587  \\
    ZPT   & 0.8909 $\pm$ 1.3278  \\
    Ours  & \textbf{0.2532 $\pm$ 0.3702} \\
    
    \bottomrule
\end{tabular}\label{tab:inv_rig}
\end{table}

\subsection{Ablation study}\label{sec:ablation}

We conducted an ablation study to demonstrate the effectiveness of our design choices by changing the network architecture. 
Four scenarios are explored: (1) without the skinning encoder, (2) without the blendshape projection loss ($L_{BP}$), (3) without the blendshape reconstruction loss ($L_{BR}$), and (4) the full model. For the case of (1), the configuration is mostly identical to those from NFR except for the output representation of the decoder, which is per-vertex displacement instead of per-triangle Jacobian. 
As the skinning block produces the localized expression code given the global code and per-vertex skinning feature, we did not experiment with the case of using the skinning block alone without the skinning encoder. 
The models were trained using the ICT and Multiface datasets, and MSE between the ground truth and the generated vertices was measured to evaluate the expression quality. 
The results are shown in Table~\ref{tab:ablation} \rev{and Table~\ref{tab:ablation_inv_rig}}.

The addition of the skinning encoder significantly improved the performance across all the data and the addition of the skinning block further enhanced the performance on the Multiface data. This indicates that the localized expression code helps the network to find the correlation between the local deformation and the global code. 
The combined use of $L_{BP}$ and $L_{BR}$ improved expression quality and led to accurate and stable inverse rigging performance, demonstrating the critical role that these losses play in refining the network's overall ability to replicate facial expressions.

\begin{table}[h]
\caption[Ablation study that measures the expression quality]{
    Ablation study that measures the expression quality. The skinning encoder is denoted as ‘SE’ and the best score is indicated in \textbf{bold}.
}
\centering
\begin{tabular}{lcc}
    \toprule 
    \multirow{2}{*}{Method}   & \multicolumn{2}{c}{MSE $\downarrow (\times 10^{-4}mm)$ }   \\ \cline{2-3}
                              & ICT      & Multiface \\ 
    \cline{1-3}

    w/o SE.  & 0.4663   & 4.9857 \\
    w/o $L_{BP}$  & 0.3771   & 4.8341 \\
    w/o $L_{BR}$  & 0.3462   & 4.5337 \\
    Full model & \textbf{0.2480}   & \textbf{4.2856} \\
    \bottomrule
\end{tabular}\label{tab:ablation}
\vspace{-1mm}
\end{table}

\begin{table}[h]
\caption[Ablation study that measures the inverse rigging quality.]{
    Ablation study that measures the inverse rigging quality. The skinning encoder is denoted as ‘SE’ and the best score is indicated in \textbf{bold}.
}
\centering
\begin{tabular}{lc}
    \toprule 
    \rev{Segments}   & MSE $\downarrow (\times 10^{-4}mm)$  \\
    \midrule
        
    w/o SE.       & 0.4081 $\pm$ 0.4776 \\
    w/o $L_{BP}$  & 0.3064 $\pm$ 0.4064 \\
    w/o $L_{BR}$  & 0.3061 $\pm$ 0.4055 \\
    Full model    & \textbf{0.2532} $\pm$ \textbf{0.3702}  \\
    \bottomrule
\vspace{-2mm}
\end{tabular}\label{tab:ablation_inv_rig}
\end{table}

\subsection{Segmetation variations}
We created a segmentation map based on the ICT model, drawing from facial muscle groups~\cite{winslow2015classic}. To determine the optimal number of segments, we experimented with several configurations. Specifically, we created four variations with segment counts of $6$, $14$, $20$, and $24$, as illustrated in Figure~\ref{fig:segment_variation} and trained our model. The results for expression quality and inverse rigging quality are presented in Table~\ref{tab:segment} and Table~\ref{tab:segment_invrig}, respectively. For the inverse rigging, we predicted the global expression code from the given ICT mesh and applied it to the basis of the ICT blendshape to reconstruct the face mesh and computed MSE with the ground truth.

\begin{figure*}[t]
\begin{center}
\includegraphics[width=\linewidth]{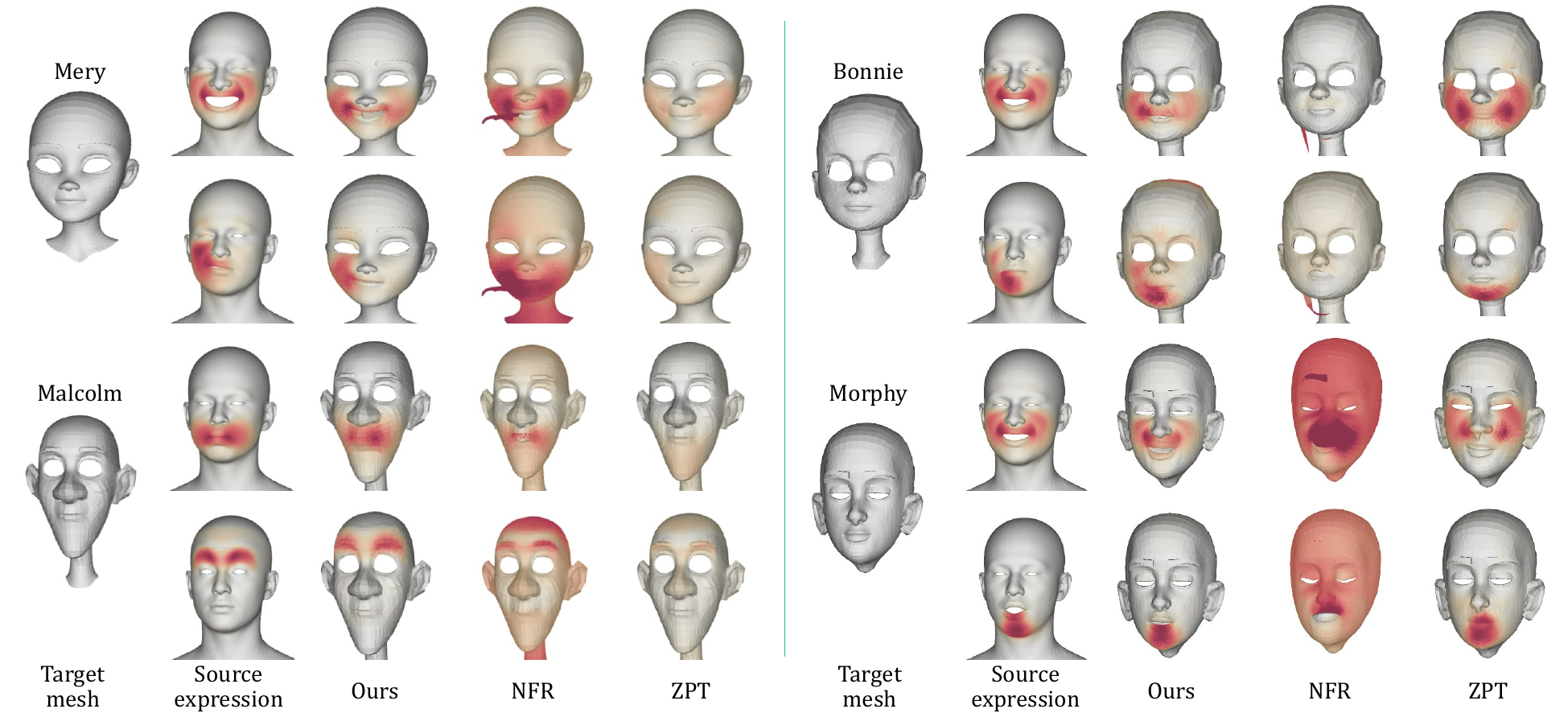}
\end{center}
\caption[Visualization of expression cloning on stylized face meshes.]{
    Visualization of expression cloning on stylized face meshes. Our method can animate arbitrary face meshes based on the ICT blendshape coefficient. The similarity of the facial movement between the source expression and ours can be observed. The MSE is colored using a yellow-orange-red color map (YlOrRd).
}
\vspace{-2mm}
\label{fig:application_unseen}
\end{figure*}

\begin{table}[h]
\caption[Expression quality with segment variation]{
    Expression quality with segment variation. The quality of expression measured from the model trained with various numbers of segmentations. 
}
\vspace{-1mm}
\centering
\begin{tabular}{ccc}
\toprule 
\multirow{2}{*}{Segments}   & \multicolumn{2}{c}{MSE $\downarrow (\times 10^{-4}mm)$ }   \\ \cline{2-3}
                          & ICT      & Multiface \\ 
\cline{1-3}


6  & 0.2538 & 4.0004 \\ 
14 & 0.2623 & 4.6797 \\
20 & 0.2480 & 4.2856 \\
24 & 0.2545 & 3.9126 \\

\bottomrule
\end{tabular}\label{tab:segment}
\vspace{-1mm}
\end{table}

\begin{table}[h]
\caption[Inverse rigging quality with segment variation]{
    Inverse rigging quality with segment variation. The inverse rigging results from the model trained with various numbers of segmentations. 
}
\vspace{-1mm}
\centering
\begin{tabular}{cc}
\toprule 
Segments   & MSE $\downarrow (\times 10^{-4}mm)$  \\
\midrule

6  & 0.2661 $\pm$ 0.3894 \\ 
14 & 0.2637 $\pm$ 0.3854 \\
20 & 0.2532 $\pm$ 0.3702 \\
24 & 0.2835 $\pm$ 0.3276 \\

\bottomrule
\end{tabular}\label{tab:segment_invrig}
\vspace{-2mm}
\end{table}

The expression quality across the models trained with each segment variation showed no significant differences. However, the model trained with 20 segments produced the best results in the inverse rigging task. Because our goal is not only to achieve good retargeting quality but also to develop a global expression code that allows for intuitive editing and manipulation, we selected the model trained with 20 segments as our final model.

\subsection{Expression cloning on stylized face mesh}


In addition to ICT and Multiface, we performed an experiment using stylized face meshes, which deviate significantly from realistic human faces. We used Mery (©meryproject.com), Malcolm (©2023 AnimSchool), \rev{Bonnie (©joshsobelrigs.com) and Morphy (©joshburton.com)} which were unseen during the training. Because the direct comparison of MSE between the source and target meshes is not feasible due to their different structures, we instead highlighted the incurred deformation of the target mesh from its neural expression using the MSE values.
The results are shown in Figure~\ref{fig:application_unseen}. Artifacts are evidently created by NFR, such as the shifted eyebrow positions and rescaled mesh as shown in the fourth \rev{and ninth} column. ZPT failed to clone the expressions for all examples. Our method accurately retargeted the source expressions to the stylized face meshes, demonstrating its versatility in handling various face shapes. Additional details and visual examples of retargeting and editing can be found in the supplementary material.





\section{Limitations}

\begin{figure}[t]
\begin{center}
\includegraphics[width=.95\linewidth]{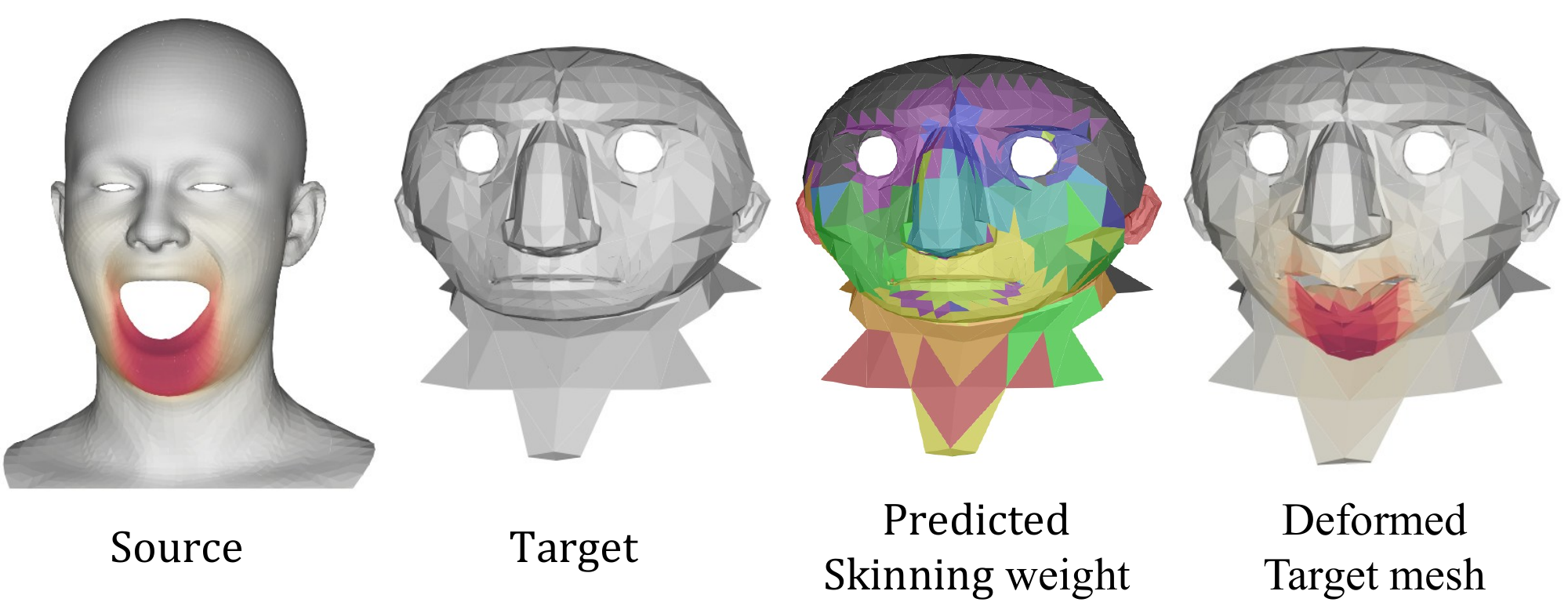}
\end{center}
\caption[Limitation]{
    The quality of expression cloning drops when the skinning encoder predicts inaccurate skinning weights from the target mesh. (©cgtrader)
}
\vspace{-2mm}
\label{fig:limitation}
\end{figure}

Although our method can effectively retarget facial expressions across different face meshes, regardless of structure, it has a few limitations. 
First, its editability is bound to the ICT blendshape which cannot handle head pose and neck movement.
Switching to a foundational model like FLAME~\cite{li2017flame}, which supports these movements, could address this limitation.
Another limitation arises from the skinning encoder because it sometimes struggles to predict accurate skinning weights on the face mesh when the scale of the given mesh differs from the learned data as shown in Figure~\ref{fig:limitation}. Furthermore, the method requires ground truth segments for training. Future work could explore unsupervised strategies for optimal segment and skinning weight prediction to move toward a more automated approach.
\rev{Additionally, although animation retargeting is possible through per-frame expression cloning, jittering sometimes occurred as shown in the supplementary video. Addressing this issue could involve integrating a temporal module, which presents a potential direction for future work.}

\section{Conclusion}

In this paper, we introduced a novel approach for facial expression cloning that combines the strengths of global and local deformation models. Our method localizes the influence of the global expression codes to improve expression fidelity. By predicting local deformation for each vertex of the target mesh, our method can be applied to meshes with arbitrary structures.
We introduced blendshape-based losses to guide our model to ensure alignment with the FACS-based blendshape which provides intuitive control over facial expressions. Throughout the experiments, we demonstrated that our approach outperforms existing methods in terms of facial expression fidelity and inverse rigging quality. 
Overall, our work highlights the importance of integrating global and local deformation strategies for accurate and flexible facial expression retargeting. 


\section*{Acknowledgements}
This work was supported by Institute of Information \& Communications Technology Planning \& Evaluation(IITP) grant funded by the Korea government(MSIT) (No.RS-2024-00439499, Generating Hyper-Realistic to Extremely-stylized Face Avatar with Varied Speech Speed and Context-based Emotional Expression)

\bibliographystyle{amsalpha-authoronly}  
\bibliography{bibo}

\end{document}